# Applications of Machine Learning for the Ratemaking in Agricultural Insurances

Luigi Biagini

*Università degli Studi della Tuscia, Italy*

**Abstract**

This paper evaluates Machine Learning (ML) in establishing ratemaking for new insurance schemes. To make the evaluation feasible, we established expected indemnities as premiums. Then, we use ML to forecast indemnities using a minimum set of variables. The analysis simulates the introduction of an income insurance scheme, the so-called Income Stabilization Tool (IST), in Italy as a case study using farm-level data from the FADN from 2008-2018. We predicted the expected IST indemnities using three ML tools, LASSO, Elastic Net, and Boosting, that perform variable selection, comparing with the Generalized Linear Model (baseline) usually adopted in insurance investigations. Furthermore, Tweedie distribution is implemented to consider the peculiarity shape of the indemnities function, characterized by zero-inflated, no-negative value, and asymmetric fat-tail. The robustness of the results was evaluated by comparing the econometric and economic performance of the models. Specifically, ML has obtained the best goodness-of-fit than baseline, using a small and stable selection of regressors and significantly reducing the gathering cost of information. However, Boosting enabled it to obtain the best economic performance, balancing the most and most minor risky subjects optimally and achieving good economic sustainability. These findings suggest how machine learning can be successfully applied in agricultural insurance.This study represents one of the first to use ML and Tweedie distribution in agricultural insurance, demonstrating its potential to overcome multiple issues.

# 1. Introduction

The agricultural sector suffers from major external shocks such as extreme weather events and market and political shocks. To handle these risks, farmers use a variety of risk management techniques, with insurance making up the lion's share (e.g., (Finger *et al.*, 2022)). Even though insurance can minimize production risks and help farmers maintain their financial stability, it does a poor job of covering a variety of hazards. Hence, farmers have a low propensity to use this tool (Glauber, 2004). With policies implemented to encourage the adoption of these instruments, many countries have attempted to address this issue, although with limited success (EC, 2009; Cole *et al.*, 2013; Banerjee *et al.*, 2019; Cai, de Janvry and Sadoulet, 2020; Feng, Du and Hennessy, 2020).

The incorrect assessment of the risk is the primary reason for the limited adoption of insurance in agriculture. In particular, this problem is caused when the premium does not correspond to the risk perception by the farmers (i.e., ratemaking). In this case, farmers may not buy the insurance and/or the insurance provision is not financially viable in the long run. Such bias in the premium specification is mainly driven by the wrong selection of variables that affect risk. Adverse selection, an issue detrimental to insurance, can be addressed through fair ratemaking (Knight and Coble, 1999; Knight *et al.*, 2010; Borman *et al.*, 2013). This creates a mismatch between the supply and demand for risk management solutions, resulting in increased neglect of insurance and, in the worst-case scenario, its failure (Babcock, Hart and Hayes, 2004; Sherrick *et al.*, 2004).

This collapse can generate cascading consequences even for other risk management tools and can be particularly paramount in the not uncommon case where they are subsidized. Taxpayers can perceive incentivized mechanisms as mismanaged, eroding trust in these instruments even more and contributing to farmers' using management risk tools even less (Park *et al.*, 2022).

Researchers and policymakers have made great efforts over the past decade to overcome these issues (Diaz-Caneja *et al.*, 2008; Meuwissen, van Asseldonk and Huirne, 2008; Chavas *et al.*, 2022); however, the challenge of the fair establishment in insurance premiums remains. Aligning premiums is often complex and/or costly, especially for new, not yet available insurance products.

One can sustain that a good forecast can be reached by employing a large set of variables (e.g. farm and farmers' characteristics). However, this approach is not exempt from critiques: first of all, it increases the expense and complexity of data gathering; second, it can be affected by collinearity and overfitting, producing poor performance.

The main goal of this study is to provide an approach to maintain high performance while using a limited number of farmers' and farms' characteristics. More specifically, we develop and test an approach based on Machine Learning (ML)(Storm, Baylis and Heckelei, 2020). ML also is expected to be a powerful approach to solving complex problems involving variable selection and collinearity (Breiman, 2001; Hastie, Tibshirani, and Friedman, 2009; Varian, 2014; Efron, 2020). Moreover, we consider a new probability density function, the Tweedie distribution. In contrast with probability density functions currently employed in the insurance field, it is very well suited to depict the indemnity distribution.

Furthermore, the combination of ML and Tweedie distribution will allow the offering of well-designed insurance contracts with premiums close to income loss perception, encouraging the participation of farmers in the insurance scheme. Our study extends earlier approaches to eliminating redundant information and selecting only essential and non-correlated variables for insurance purposes. For example, El Benni, Finger and Meuwissen (2016) used stepwise regression and genetic algorithms with significant results. Still, these algorithms are not specifically designed to reduce the collinearity problem.

We apply the here developed framework using an empirical example of the Income Stabilization Tool (IST) in Italy. This is a new income insurance solution that is not yet implemented on the market on a large scale but is supported by specific policies. Thus, this case study provides an example where an efficient ratemaking system is needed, e.g. by reducing the variables required for the ratemaking system. This is crucial because while considering numerous variables could increase the accuracy of insurance prices, doing so would imply more costs.

Our results suggest that the considered ML tools were well suited to developing insurance products, i.e., setting premiums, even though the different considered ML methods differ concerning the goodness-of-fit and the number of selected variables. Furthermore, the economic assessment showed that using ML tools could improve the sustainability of the developed insurance products by setting premiums at reasonable levels. Finally, these findings suggest that ML can be successfully applied in other fields of agricultural insurance.

This paper is organized as follows: section 2 offers a perspective on insurance ratemaking and the IST, putting our research within the literature context. The methods and data used in the analysis are discussed in section 3. Section 4 discusses the results, addressing some key econometric issues and assessing whether the use of ML estimates in setting the premium could improve the economic performance of the scheme. Finally, the final section summarizes the study findings, emphasizing the proposed approaches' advantages, methodological limitations, and potential application constraints and recommending future research areas.

## *2.* Empirical strategy

The empirical strategy was based on disentangling the aspects of insurance instruments' attractiveness. In addition, we describe the Income Stabilization Tool because it represents an optimal case study due to its characteristics. It is, in fact, a new tool, poorly implemented and with a high possibility of running into disaffection or failure. Finally, we introduce the theoretical framework and explain the theoretical framework of our analysis.

### 2.1  *The challenges for insurance attractiveness: ratemaking, fairness, affordability, and economic sustainability*

Insurance can be defined as *"the exchange of money now for money payable contingent on the occurrence of certain events"* (Arrow, 1965; Zweifel and Eisen, 2012) or *"the exchange of an uncertain loss of unknown magnitude for a small and known loss (the premium)"* (Hax, 1964; Zweifel and Eisen, 2012). The insurance contract specifies the premium ($Prem$), or the disbursement needed to buy the contract, and the indemnity *(Ind)* which is the payment that the insurer reimburses to the policyholder after a specific event occurs. Hence, the insurance contract can be described as a two-dimensional vector $s = \{s \in \mathbb{R}^2 : (Prem, Ind)\}$. The indemnity depends on the occurrence losses *(Loss)* and is defined as the $Ind = Loss - (a + d \times Loss)$. The term into brackets is the deductible level that can be defined as a fixed level $a$ plus a proportion $d$ of the loss. According to Wilson, (1977), assuming disposable incomes ($I$), the "space of insurance" contract is defined as:

$$\bar{S} = \{s \in R^2 : I - Loss - Prem + Ind \geq 0, I - Prem \geq 0\} \tag{1}$$

We assume that insurance bases the premium on ratemaking, defined as the process of establishing a connection between the level of risk and the premium to be paid (Dionne and Rothschild, 2014; Vaughan and Vaughan, 2014; McNamara and Rejda, 2016). Thus, while the exact computation of the expected indemnities is an essential step of ratemaking, a function that connects the expected indemnity with observable policyholder variables ($X_i$) - in our case, individual farms and farmers' characteristics - are required (Bernard, 2013).[1]

$$E(Ind) = f(X_i) \tag{2}$$

The insurer needs to cover expenses (and profit) to manage this instrument, including "loading costs" that are assumed to be proportional to indemnity: $\delta E(Ind_{i,t})$. Hence, the insurance premium of the individual $i$ at time $t$ ( $Prem_{i,t}$ ) is defined as the expected indemnity $E(Ind_{i,t})$ and loading costs $\delta E(Ind_{i,t})$.

$$Prem_{i,t} = E(Ind_{i,t}) + \delta E(Ind_{i,t}) \tag{3}$$

The challenge of an insurance contract is to establish a fair premium that should be high enough to remunerate the insurance but low enough to induce buyers to enrol (Finn and Lane, 1997).

In this investigation, we relaxed the assumption of mandatory subscription that the literature has assumed for IST ( f.e. Rippo and Cerroni (2022); Severini, Biagini and Finger (2019); Liesivaara and Myyrä (2016)). However, assuming voluntary subscription opens up new and complex scenarios. In particular, one can find subjects who pay too much for the risk taken and those who pay too little. As a result, we can find insureds with overstated losses subsidizing policyholders with undervalued risks. The consequence is that the less risky subjects can no longer subscribe to the insurance, leaving only the riskiest subjects, increasing the chances of failure of the insurance scheme. In contrast, an insurance scheme is perceived as "fair" when this is not the case (Babcock, Hart and Hayes, 2004; McNamara and Rejda, 2016).

Participation was also reduced when the premium was not "affordable"(Goodwin, 2001), or when the premium should be compatible with the potential insured's wealth, in our case, assuming as the proxy the disposable income (Wilson, 1977; Zhang and Palma, 2020). Nothing can function if the premium is incompatible with the policyholder's expense capacity, regardless of how fair the insurance scheme is. In summary, we assumed that "fairness" and "affordability" are sufficient and required conditions to have a high level of participation in an insurance scheme.

---

[1]Although it would be desirable for this type of analysis to include historical trends and cycles, the lack of a dataset with a long time series did not allow this type of approach to be used (Goodwin and Mahul, 2004)

These aspects mainly affect policyholders' behaviours, but the insurer can assess other aspects. In particular, according to Goodwin & Mahul, (2004), an insurer is interested in evaluating the new 'insurance's overall financial results of the insurance scheme, assessing the difference between premiums and indemnities over a suitable number of years ($\Pi_{t=0}^{T}$):

$$\Pi_{t=0}^{T} = \sum_{t=0}^{T} \sum_{i=1}^{N} \left( Prem_{i,t} - Ind_{i,t} \right) \qquad (4)$$

Where, $\Pi_{t=0}^{T}$ represents the multiannual balance sheet that can be $\Pi_{t=0}^{T} \geq 0$; otherwise, insurance is not convenient, and it will exit the market.

Finally, insurers generally wish to avoid large fluctuations in their annual financial results. This is because significant unexpected losses can threaten the 'insurer's financial position (i.e., solvency) if no appropriate strategies (e.g., financial reserves or reinsurance contracts) are installed. Therefore, attention should also be paid to the fluctuations in the annual financial results.

The above-mentioned aspects of accessibility and fairness for the producer may conflict with the desire for low fluctuation on the insurance side. In particular, the trade-off between the insurance profit, which cannot be too high, and the insured, who do not pay too much in the case of a high premium compared to their own risk (caused by high loading costs), is vital. An excessive unbalance in this trade-off can lead to the abandonment of the insurance tools by policyholders towards other forms of risk management and the consequent failure of the insurance scheme. Therefore, a multifaced evaluation is necessary to evaluate these aspects simultaneously to avoid failure and increase the insurance scheme's attractiveness.

## 2.2  The choice of the case study: Income Stabilization Tool, functioning, and implications

To overcome issues in the insurance scheme, we need to study a realistic case. Therefore, we focused on insurance or similar tools that cover income failure and volatility risk.

The EU knows several instruments (EC, 2001, 2017b; Cafiero *et al.*, 2007; Diaz-Caneja *et al.*, 2008; Meuwissen, van Asseldonk and Huirne, 2008; Meuwissen, Assefa and van Asseldonk, 2013), including whole farm income insurance schemes such as the Income Stabilization Tool – IST – (EC, 2010, 2011b, 2011a, 2013a, 2013b).

IST represents an excellent case study because it represents a novel (similar) insurance tool characterized by a low participation rate and a high potential benefit for agriculture (see f.e., (Capitanio, Adinolfi and Pasquale, 2016; Severini, Biagini and Finger, 2017; Severini, Tantari and Di Tommaso, 2017; Trestini *et al.*, 2018; Giampietri, Yu and Trestini, 2020; Chavas *et al.*, 2022; Rippo and Cerroni, 2022)).

IST partially compensates for losses incurred by participating farmers. This scheme is managed by a mutual fund owned by associated farmers; hence, it is designed to have zero extra profits. However, EU regulations do not specify how farmers should contribute to the mutual fund. Therefore, this study uses a market-based approach considering the IST is similar to an insurance scheme.

According to the EU regulation, the trigger level is set at 70% of the previous three-year income arithmetic average or the Olympic average of the last five years [2]. The maximum indemnity level is 70% of the trigger level and the current income difference: this strategy is used to reduce moral hazard issues (Goodwin and Mahul, 2004; Cordier and Santeramo, 2020; Wu, Goodwin and Coble, 2020). Following Finger and El Benni, (2014), we have chosen the arithmetic average of the past three years of income to identify the trigger level.

IST provides compensation to farmers who experience an income drop of more than 30% compared with the expected income level (European Parliament, 2016; EC, 2017a; Meuwissen, Mey and van Asseldonk, 2018) regardless of the causes of such drop. The indemnity paid for the $i_{th}$ farm at the $t_{th}$ year, is defined as follows:

$$Ind_{i,t} = \begin{pmatrix} 0 & if & I_{i,t} \geq I_{R\,i,t} \\ b\left[ E(I)_{i,t} - I_{i,t} \right] & if & I_{i,t} < I_{R\,i,t} \end{pmatrix} \qquad (5)$$

where:

---

[2] The use of Olympic means is not considered in this study to avoid reducing the sample too much. Consider that the Olympic average requires the availability of the last five years, as opposed to the three years used for the arithmetic average. Even using the arithmetic mean, the sample is necessarily smaller than the original because the farms with three consecutive years of observations are not always available. We can use a value to fill the gap of one year, i.e., the average of two years, but this exercise is beyond the scope of this article.

$I_{i,t}$     is the realised income.

$E(I)_{i,t}$       is the expected income based on the average of the realised $I_{i,t}$ of the previous three years (see Finger and El Benni 2014b for discussions) as $E(I)_{i,t} = \frac{1}{3}\sum_{t-1}^{t-3} I_{i,t}$;

$I_{R\,i,t}$ is the trigger level defined as: $I_{R\,i,t} = a\,E(I)_t$ where parameter $a$ EU regulations set at 0.7.

The EC regulation also sets deductible as 30% and $b = 100\% - 30\% = 70\%$

Consequently, the space of indemnity $\overline{Ind}$ of IST is defined as

$$\overline{Ind}: \{Ind \in \mathbb{R}: Ind(0 \cup min_{Ind}, +\infty)\} \qquad (6)$$

Where $min_{Ind}$ is the minimum value of $b\left[E(I)_{i,t} - I_{i,t}\right]$ among all the policyholders.

## 2.3  Assumptions

IST is managed by a mutual fund (MF) that sets premiums and pays indemnities. In this study, we approximate MF as an entity that operates in the market without the objective of maximizing its profit but to maintain an equilibrium between revenues (premiums) and costs (paid indemnities in our analysis). Furthermore, we assume the IST is managed by only one national mutual fund without losing generality.

Our study relaxes the mandatory nature of IST: the policyholder can choose to either adopt this tool or not. This condition allows us to have a near-reality investigation. We assume that farmers' subscription choice derives from the contemporaneous examination of two aspects: "affordability" (evaluation of compatibility between premium and expense capability) and "fairness." which derives from assessing the insurance scheme's difference between premiums and indemnities (the net premium).

The optimal condition should establish a premium equal to the expected losses. The insurer, unable to directly verify the expected level of losses and the indemnities by the individual farmer, uses the farm's characteristics and the farmer's to make a solid forecast of the expected losses.

The timing of farmer behaviour can be described considering two periods $t$ and $t_{-1}$. The insurer defines the premium after gathering farm information. Farm information derives from a balance sheet that refers to $t_{-1}$ period. Consequently, the expected indemnity for time $t$ can be estimated using only $t_{-1}$ variables. On the other hand, the farmer decides on buying insurance or not, taking into account her/his willingness to pay by relying on income at $t_{-1}$. In sum, the premium and the decision to participate in IST depend on the information at $t_{-1}$.[3]

Even in a high-stress situation, the farmer can change his behaviour. Therefore, the variables that affect income and risk perception in one year could be significantly different from those in another. The evaluation using the year-to-year approach allows us to consider this aspect.

Moreover, no information is available for loading costs, usually used in insurance. This cost item, in each case, would only lead to a proportional increase in the premium, affecting only the willingness to pay and not the ratemaking. Consequently, the assumption that the loading cost is equal to zero does not change the result of our investigation.

Furthermore, it is assumed that the MF proposes a homogeneous contract to all farmers, with the level of premiums as the only parameter that changes from one contract to another.

Lastly, because it is impossible to capture the impact of 'farmers' insurance behaviour in new insurance products (Goodwin and Mahul, 2004), farmers are assumed not to adjust their behaviour after participation. This further reduced the impact of moral hazard.

## 3. Data and estimation strategy

We employed the panel dataset from the Italian Farm Accountancy Data Network FADN in 2008–2018 of 118,748 individual farm-level observations. All economic values were deflated using the Eurostat Harmonized Index of Consumer Prices[4].

---

[3] Note that using the covariates at the time $t_{-1}$ (Werner, Modlin and Watson, 2016), solves reverse causality and simultanuity bias (Bellemare, Masaki and Pepinsky, 2015; Fajgelbaum, Schaal and Taschereau-Dumouchel, 2017).

[4]Available at: https://ec.europa.eu/eurostat/web/hicp/data/database

The estimation strategy was developed along the steps represented in Figure 1.

In the first step, an ex-ante numerical simulation of the IST implementation in Italy is developed. The analysis is based on farm net added value-added for the computation of IST indemnity, according to the decisions of the Italian government (ISMEA, 2015; Mipaaf, 2017). This was given by farm revenues and public payments (e.g., CAP direct payments) minus costs for external inputs. The simulation allows for obtaining the potential distribution and levels of farm-level indemnities.

In the second step, the expected indemnities (later on used as premiums) were estimated based on a large set of variables related to farm and farmers' characteristics: $E(Ind_{i,t}) = f(X_{i,t-1})$.

In the last step, we assessed the econometric performance (Step 3a) and the economic performance (Step 3b) of the considered models. The latter was based on the assumption that the expected indemnities were used to establish the premiums.

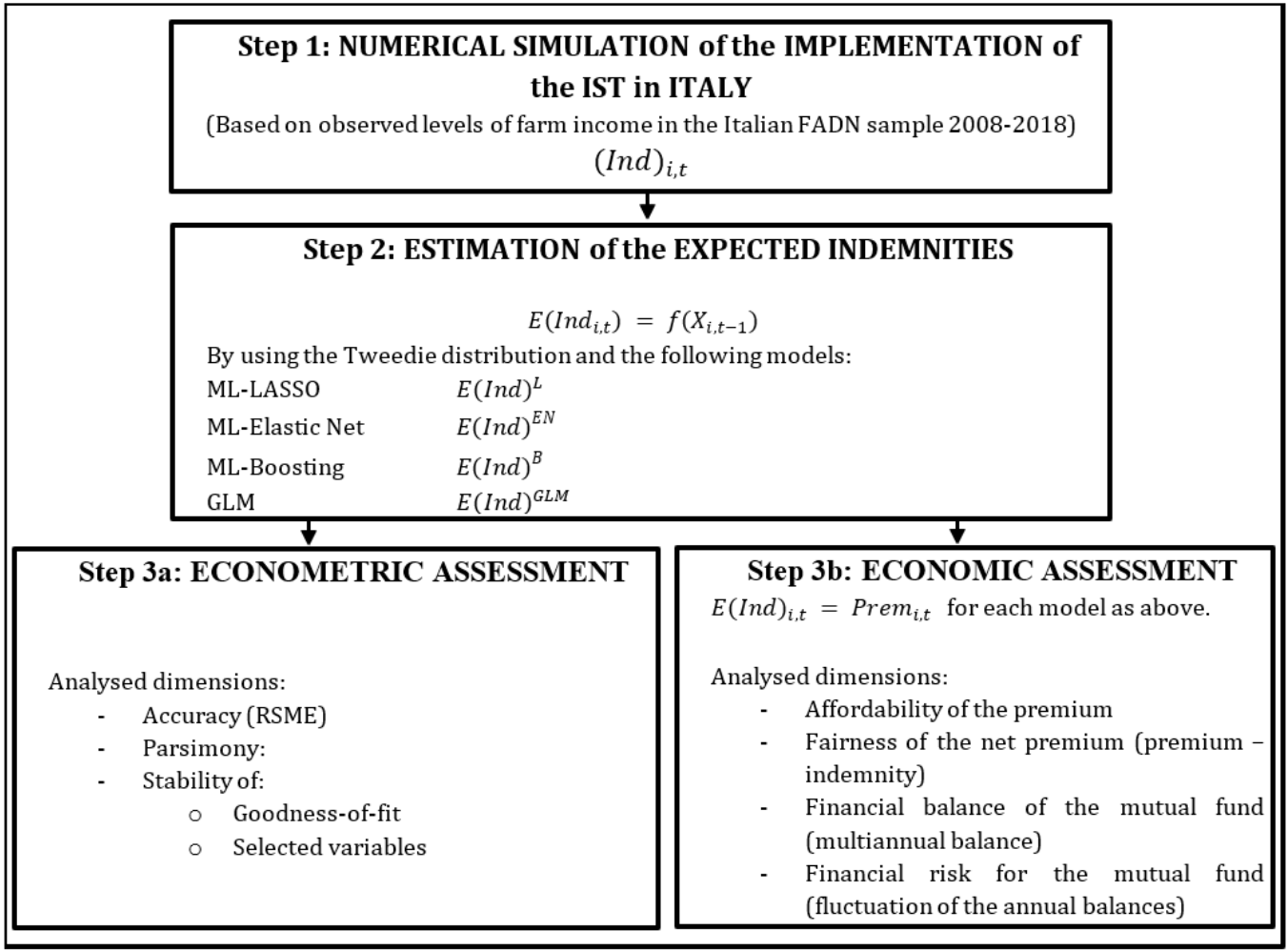

**Figure 1 Graphical representation of the steps of the analysis.**

## 3.1 Simulation of IST

In the first step of the analysis, we simulated the expected IST indemnities following the approach of Goodwin and Mahul (2004). The value-added of the three years before the time was used to find the trigger level. As the IST is not yet implemented in Italy, the trigger level is not available for the first three years, and consequently, the period from 2008 to 2010 cannot be used for estimation. Furthermore, only farms with positive reference incomes are maintained because the EU regulations do not explain how the IST should consider negative incomes (0.48% of the total sample). Finally, 47.898 observations remain for the analysis.

The indemnity distribution obtained by this simulation is peculiar, being zero-inflated, right-skewed, and thick-tailed (Werner, Modlin, and Watson 2016), with a minimum indemnity value of € 236.40 ( Table I and Figure 2).

|  | Nr Obs | Mean | Standard deviation | Median | Min | Max | Skewness | Kurtosis |
|---|---|---|---|---|---|---|---|---|
| Sample | 47898 | 7154.75 | 46644.17 | 0.00 | 0.00 | 3111487 | 26.83 | 1094.01 |
| Only with indemnity >0 | 10592 | 32354.45 | 94994.44 | 12377.22 | 236.40 | 3111487 | 13.55 | 270.56 |

**Table I - Descriptive statistics of the farm-level indemnities as simulated in the whole farm sample in the considered years (Euro/farm).**

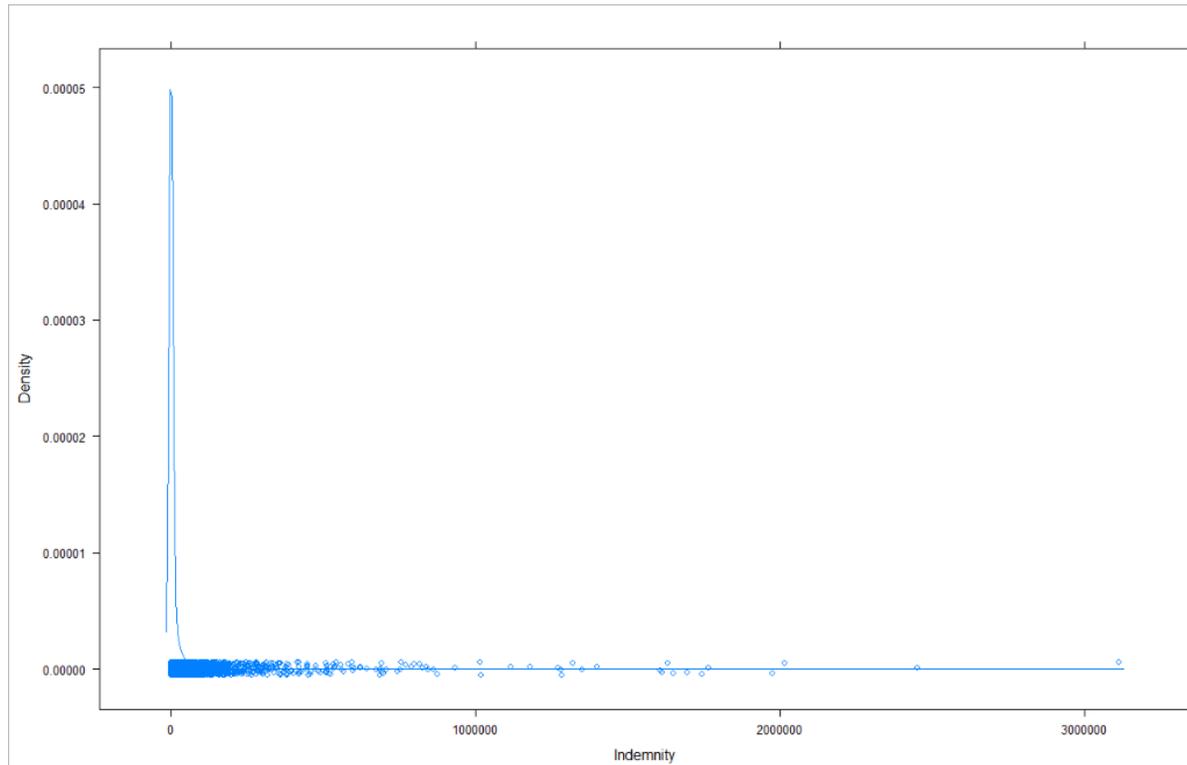

**Figure 2 – Density plot of indemnity (On the abscissas, the values of indemnities are in Euro)**

At this point, defining which variables can affect the indemnity level was necessary. The literature confirms that several factors can potentially affect the income downside risk and, consequently, IST indemnity (El Benni, Finger and Meuwissen, 2016). Topographic, climatic, and socio-economic conditions affect the relative farm's profitability and the availability and characteristics of production factors and functions. Farmers in the mountain regions face higher (relative) income variability than those in other regions (El Benni, Finger and Mann, 2012; Severini, Tantari and Di Tommaso, 2016).

Farm characteristics such as farm size, production characteristics, and financial features have also been found to be relevant (e.g., (Mishra and El-Osta, 2001; Yee, Ahearn and Huffman, 2004)). Large farms were found to better manage extreme events compared to small farms (El Benni, Finger and Mann, 2012). Different farming types face income variability levels (Severini, Tantari and Di Tommaso, 2017). Similarly, production intensity affects production risk (see, e.g., (Busato *et al.*, 2000; Busato, Trachsel and Blum, 2000; Schläpfer *et al.*, 2002; McBride and Greene, 2009; Gardebroek, Chavez and Lansink, 2010; D'Antoni and Mishra, 2012)). On-farm diversification decreases downside income risk (e.g., (Di Falco *et al.*, 2010)). Similar considerations apply to farm financial characteristics, such as cost flexibility, liquidity availability, credit use, and farm insurance (Jodha, 1981; Robison and Barry, 1987; El Benni *et al.*, 2012; Hardaker *et al.*, 2015). Farmers' characteristics, such as gender and age, influence risk preferences and perceptions (Hartog, Ferrer-i-Carbonell and Jonker, 2002; Bruce J. Sherrick *et al.*, 2004; Menapace, Colson and Raffaelli, 2013). Agricultural policies can also affect farm income risk. In particular, in the EU, a large share of farm income is generated by CAP payments that reduce income variability, making it a relatively stable income source (Finger and Lehmann, 2012; El Benni, Finger and Meuwissen, 2016; Severini, Tantari and Di Tommaso, 2016). Finally, income volatility experienced in the past could be used to predict the size of the indemnities.

In our investigation, the variables used can be generally classified into general, production, financial, and policy farm characteristics. For crucial variables, such as income, the model includes the level and the variability over the previous three years as standard deviation[5]. Furthermore, we include dummies to consider gender, age, organic production, property type, altimetric zones and the regional factors' differences. Unfortunately, it is impossible to account for off-farm incomes because the FADN does not provide this information. Table II reports the general statistics for the sample used to estimate the expected IST indemnities[6].

| Description | Code | Value (standard deviation) |
|---|---|---|
| *Farm general characteristics* | | |
| Utilized agricultural land [Ha] | LND | 33.29 (57.20) |
| Livestock Units [LU] | LU | 49.35 (363.39) |
| Labour input [AWU] | LAB | 1.96 (2.47) |
| Total revenues [€] | REV | 141011.40 (433029.20) |
| *Farm production characteristics:* | | |
| Livestock intensity [LU/Ha] | LU I | 3.64 (105.05) |
| Machinery intensity [kWh/Ha] | MACHIN | 16.00 (30.69) |
| Labor per UAA [UL/ha] | LAB | 0.28 (0.86) |
| Land productivity [€/ha] | VA | 7769.13 (52011.44) |
| Specialization [Herfindhal Index] | H Index | 0.67 (0.25) |
| Other Gainful Activities [%] | OGA | 0.36 (8.61) |
| *Farm financial characteristics:* | | |
| Fixed cost [€/ha] | FXCOST | 1642.36 (7405.89) |
| Current over total costs [%] | CURCOST | 0.65 (0.19) |
| Labor over total costs [%] | LBRCOST | 0.20 (0.15) |
| Insurance premia over total costs [%] | INSURE | 0.02 (0.04) |
| Relative amount of fixed capital [%] | FXK | 0.57 (0.93) |
| Relative amount of debts [%] | DEBT | 0.02 (1.21) |
| Relative amount of net worth [%] | NETK | 0.98 (1.21) |
| *Farm policies (CAP)* | | |
| Decoupled Direct Payments [€] | DDP | 47.31 (240.72) |
| Coupled Direct Payments [€] | CDP | 48.66 (332.34) |
| Rural Development Policies- Agroenvironmental [€] | RDP_AES | 56.73 (193.89) |
| Rural Development Policies - Less Favoured Area [€] | RDP_LFA | 35.34 (189.79) |
| Rural Development Policies for Investments [€] | RDP_INV | 30.15 (994.90) |
| *Other farm characteristics as number of observations:* | | |
| Number of farm | N_FARMS | 47898 |
| Sole proprietor farm | INDIV | 41742 |
| Gender of the holder | MALE | 38090 |
| Young holder | YOUNG | 4699 |
| Organic farms | ORGAN | 5455 |
| Plain regions | PLAIN | 10967 |
| Hill regions | HILL | 21028 |
| Mountain regions | MOUNT | 15903 |

**Table II - Descriptive statistics of the farms' characteristic**

## 3.2 Estimation of the expected indemnities

As a second step, the expected indemnities and, thus, premiums were regressed based on a large set of variables related to farm and farmers' characteristics in the previous year: $E(Ind_{i,t}) = f(X_{i,t-1})$. This evaluation faced two main issues: the peculiarity of indemnities distribution and the need for variable selection variables.

---

[5]A longer interval to calculate the standard deviation was not considered because this would have resulted in an excessive reduction of the available sample size. We also do not include the square or higher powers. The simulation with this type of transformation (also with polynomial form) was taken into consideration during the various settings for the implementation of the models. The ML tools used in this study automatically choose one and only one type of transformation for each variable, considering that any exponentiation is collinear to the variable of order one (in level). It should also be considered that boosting can also evaluate a transformation of a variable. Results obtained considering the squared variables can be obtained upon request.

[6] The full list of these variables can be found in Section 5 of the Supplementary material.

The set of possible indemnities (i.e., the space of the indemnity) follows peculiar rules (see table Table I and Figure 2): i) a large number of farms has no indemnities generating a zero-inflated distribution, ii) which is very far from a normal distribution (with skewed and leptokurtosis), iii) discontinuity. Table III shows how, also in our case, the space of indemnities has a particular probability density function that poses a challenge for a maximum likelihood estimation (MLE) usually used in insurance to forecast indemnities (Goodwin and Mahul, 2004).

We compared different probability density functions and concluded that the Tweedie (Jørgensen, 1987; Jørgensen and Paes De Souza, 1994)distribution accounts at best for all characteristics of the space of indemnities (Table III).

| Probability Density Function | Support | Zero & $R^+$ | Zero Inflated | Tick Tailed | Asymmetric Distribution | Discontinuity Function (Deductible) | Continuous after Min Value of Indemnity |
|---|---|---|---|---|---|---|---|
| Normal | $-\infty \leq x \leq +\infty$ | No | Partially | No | No | No | Yes |
| Poisson | $\mathbb{N}$ | Yes | Yes | No | Yes | Yes | No |
| Negative Binomial | $\mathbb{N}$ | Yes | Yes | Yes | Yes | Yes | No |
| Inverse-Gaussian | $x \in (0, \infty)$ | Yes | Yes | Yes | Yes | No | Yes |
| Gamma | $x \in (0, \infty)$ | Yes | Yes | Yes | Yes | No | Yes |
| Tweedie | $x \in (0 \cup_{min} x^+, \infty))$ | Yes | Yes | Yes | Yes | Yes | Yes |

**Table III - Characteristics of Probability Density Functions that are potential candidates for representing the space of IST indemnities**

Tweedie distributions are Poisson-Tweedie mixtures belonging to the Exponential Dispersion Models and can overcome the El Benni, Finger and Meuwissen (2016) model that faced this issue using a double-hurdle model with a Box-Cox transformation. One note is that the Box-Cox transformation is unsuitable with zero lower bound and can introduce bias in the estimation (Nelson and Granger, 1979).

Variable selection is crucial because many factors influence the expected indemnities, as confirmed by a large body of literature. However, while using many variables as regressors may improve forecast performance, it has two drawbacks: it may introduce unnecessary distortions (which should be corrected) and may generate relevant costs that translate into higher premiums for farmers to collect, tidy, and manipulate the data (Spence, 1973; Rothschild and Stiglitz, 1976; Arrow and Arrow, 1994; Varian, 2014).

In this study, we attempt to overcome the issue of the trade-off between the number of variables and the forecast performance (El Benni, Finger and Meuwissen, 2016) using only ML tools that perform variable selection and, in particular[7]: Least Absolute Shrinkage and Selection Operator -LASSO- (Tibshirani, 1996), Elastic Net -EN- (Zou and Hastie, 2005) and Boosting (Friedman, 2001) (Table IV).

---

[7]A discussion about Machine Learning and how it differs from classical inference can be found in Sections 1 and 2 of the Supplementary material. ML does not provide confidence intervals. Athey and Imbens, (2019) highlight that it is not always necessary to estimate confidence intervals, especially if this imply sacrificing other analysis objectives.

| Model Acronym | Description | Type of Estimation | Main Reference |
|---|---|---|---|
| GLM | Generalized Linear Model with all the regressors | Maximum Likelihood Estimation | Jorgensen and Paes De Souza (1994) |
| LASSO | Regularized method with fixed shrinkage parameters $\alpha = 1$ | Machine Learning | Tibshirani (1996) |
| EN | Regularized method with shrinkage parameters variables $\alpha = 0.1 \div 0.9$ | Machine Learning | Zou and Hastie (2005) |
| Boosting | Algorithm which converts weak learners to strong learners | Machine Learning | Friedman (2011) |

**Table IV- Main features of the considered GLM and Machine Learning models**

Finally, we briefly describe the estimation procedure. First of all, the parameter ($p_{opt}^*$) that conditions the shape of the Tweedie distribution is established using the R-package "TWEEDIE" (Dunn, 2017). In this step, we removed outliers through the approach of Chatterjee and Hadi (2009). The Tweedie function $p_{opt}^*$, which was used in this phase, serves as the foundation for the GLM and ML models.

Specific algorithms and packages are applied for each ML tool. In particular, for Elastic Net and LASSO, the "HDtweedie" (Qian, Yang and Zou, 2013), and for Boosting, the "TDboost" (Yang, Qian and Zou, 2016) R packages are used.

We relied on the "out-of-sample" procedure (Tashman, 2000) that evaluates the model's robustness to prevent overfitting cases. This procedure relies on ten random subsets per year, with a training set containing 75% of the observations for every random group. For the test set, all observations of the following year were used.

The Year-to-Year approach with the out-of-sample procedure allows the discovery of "transitory patterns" (Breiman, 2001), i.e., models that are valid only for a particular subset of observations but not in general (some authors define this as a local solution vs a global solution). Theoretically, the most robust models should tend to have only one pattern.

Finally, grouped regression is used to handle variable selection properly in the case of dummy variables assuming multiple discrete values (i.e., levels) following the approach by Qian, Yang and Zou (2016) [8].

## 3.3 Econometric and economic evaluations

The econometric evaluation is based on assessing goodness-of-fit outcomes, their variability, and the number and stability of the selected variables.

The goodness-of-fit concerns the accuracy of the model in predicting the indemnities and, consequently, premiums. A higher goodness-of-fit value indicates a smaller gap between expected indemnities and premiums, as well as less dissatisfaction with the insurance scheme. We employed Root Mean Squared Errors (RMSE) as a standard in the ML to assess the goodness-of-fit (Hastie, Tibshirani, and Friedman 2009; P. Murphy 1991; Van der Paal 2014). RMSE considers non-linearity between the expected and simulated compensation values to be typically present in insurance schemes and, simultaneously, allows overcoming the trade-off between bias and variance commonly affecting ML tools. Note that less complex (generally also less flexible) models tend to have significant prediction bias but low variation (contrary to highly complex models). RMSE is the sum of the root of the difference, powered at squared, between estimation and value squared and can be thought of as the variance of squared bias. This metric is a robust model selection criterion, considering the double aspect evaluated simultaneously (Storm, Baylis and Heckelei, 2020). Next, the results of the out-of-sample procedure are evaluated to assess the goodness-of-fit variability. Out-of-sample can highlight overfitting phenomena: i.e., the model has excellent forecasting performance in the subset used to obtain the parameters, but, using other sets of data, it performs poorly. This aspect is evaluated by examining the variability of the goodness-of-fit in ten out-of-sample procedures for each year (i.e., 70 procedures for each year). In general, the model with limited variability is preferred.

One key point of this research is variable selection because dealing with a low number of variables (i.e., high selection capability) is a desired property from the insurance and governmental point of view. Nevertheless, this aspect cannot be disconnected from the stability of the number of selected regressors. If the variables selected differ across out-of-samples, the model results are unstable and may not be suitable for indemnity estimation because the insurance cannot correctly select the information to collect. Therefore, a model with a small, stable set of variables is preferred.

The economic evaluation is assessed by analysing the results obtained using the model predictions for ratemaking (i.e., to define the premium levels). This is done considering three aspects: affordability and fairness of the ratemaking and overall economic performances of the scheme.

---

[8]This aspect is discussed in Section 4 of the Supplementary material

Affordability refers to the compatibility between the premium and expense capability of the insured. This latter is assessed using the ratio of *premium over farm value-added*. These individual farm ratios are compared with the observed ratios of *farm expenses for insurance premiums,* referring to already available insurance. A sensitivity analysis is also carried out as it is impossible to precisely identify the level of premium that can be considered not affordable. Therefore, we consider four thresholds referring to the 50th, 90th, 95th, and 99th percentiles of the observed ratio in the whole sample of farms.

For the evaluation of the "fairness" of the ratemakings, it is necessary to examine the balance of insurance by comparing the "net premium" (difference between premium and indemnity) of all participants. The measure of "fairness" is given by the number of insurers that significantly imbalanced the distribution of net premiums. The greater the disparity, the less fair the ratemaking is.

Finally, we consider the overall financial results of the MF managing the IST, referring to the entire period (multiannual) and for each year (annual). The multiannual evaluation allowed us to assess how well the different ratemaking approaches ensure the financial sustainability of the IST over the whole period. The analysis of the annual balances enables the assessment of the fluctuations of the economic results of the MF over considered years. This allows for spotting significant and unexpected losses (due to high indemnities paid) that can threaten the 'insurer's solvency if no appropriate strategies are pursued.

# 4. Results

## 4.1 Econometric evaluation

The econometric evaluation is based on goodness-of-fit outcomes and the number and stability of the selected variables.

### 4.1.1 *Evaluation of goodness-of-fit: outcomes and their variability*

The mean value of RMSE shows how GLM performed very poorly compared with ML tools (Table V), having a higher level of RMSE. Furthermore, the analysis of standard deviations suggests GLM does not guarantee a stable performance in the subset, suggesting the presence of overfitting. Conversely, ML tools demonstrate high performance, particularly for LASSO and Boosting. Finally, the stability of these results is confirmed based on the comparison of the results for all years.

| Year Model | 2012 | 2013 | 2014 | 2015 | 2016 | 2017 | 2018 |
|---|---|---|---|---|---|---|---|
| GLM | 15.90 | 16.67 | 13.32 | 22.68 | 78.90 | 57.63 | 102.81 |
| | *(16.36)* | *(17.18)* | *(13.77)* | *(23.18)* | *(79.20)* | *(58.11)* | *(103.30)* |
| LASSO | 5.74 | 7.59 | 4.73 | 8.01 | 4.78 | 5.97 | 4.74 |
| | *(6.19)* | *(8.09)* | *(3.81)* | *(8.35)* | *(4.89)* | *(6.38)* | *(4.83)* |
| EN | 8.28 | 7.54 | 4.81 | 9.58 | 6.91 | 6.81 | 25.76 |
| | *(8.77)* | *(8.04)* | *(4.32)* | *(10.08)* | *(7.40)* | *(7.30)* | *(26.26)* |
| Boosting | 4.70 | 4.77 | 4.72 | 4.57 | 4.55 | 4.51 | 4.52 |
| | *(3.42)* | *(3.41)* | *(3.48)* | *(3.55)* | *(3.02)* | *(2.94)* | *(3.23)* |

**Table V- Goodness of Fit: mean of the Log RMSE for all years and ten test samples per year (standard deviations are in italics in parentheses)**

### *4.1.2    Number and stability of the selected variables*

The average number of variables selected is reported in Table VI. While GLM uses all 129 regressors, the considered ML performs variable selection. EN and LASSO are the most parsimonious models and select from 4 to 15 regressors and from 10 to 24 regressors, respectively. In contrast, Boosting is less parsimonious, selecting between 42 and 54 variables.

| Year Model | 2012 | 2013 | 2014 | 2015 | 2016 | 2017 | 2018 |
|---|---|---|---|---|---|---|---|
| GLM | 129 | 129 | 129 | 129 | 129 | 129 | 129 |
| LASSO | 9 | 4 | 4 | 15 | 7 | 4 | 7 |
| EN | 19 | 12 | 15 | 24 | 11 | 10 | 19 |
| Boosting | 52 | 52 | 54 | 49 | 54 | 42 | 45 |

**Table VI- Number of variables selected in mean per year in all models**

At this point, assessing the stability of the selected variables is helpful since instability implies that the model cannot be based on a stable set of variables (Figure 4).

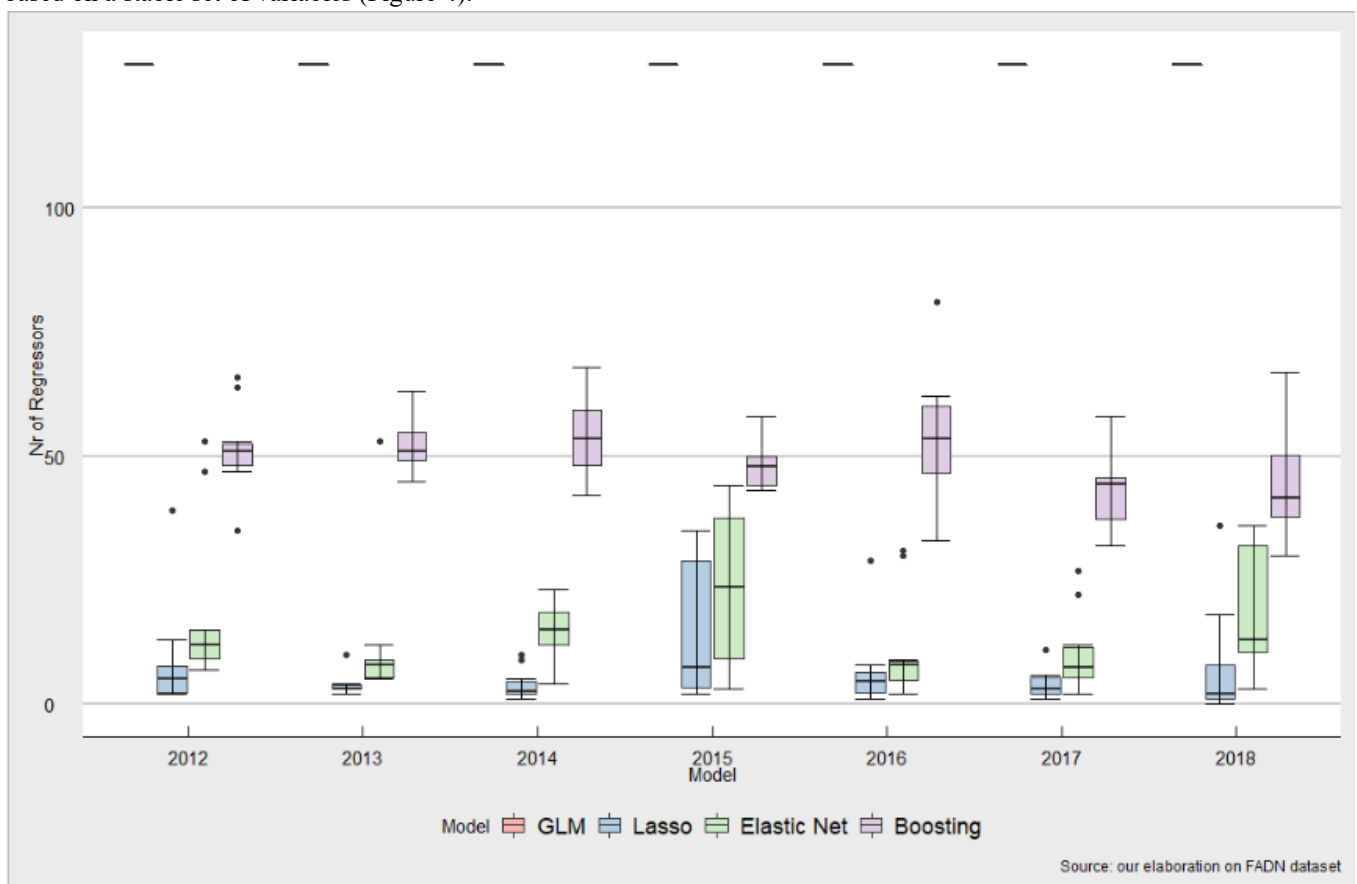

**Figure 3 – Boxplot of the number of variables selected in all year and tests. Results for the GLM and ML models.**

In particular, Boosting selects a relatively stable number of regressors across all years, indicating that its predictions are consistent. However, looking only at the number of selected regressors may not be sufficient to support this finding, as the selected variables may vary. To address this issue, we assess whether the selected regressors are the same in the different considered cases (resamples and years), calculating each regressor's selection frequency. This allows for assessing ML stability and identifying the variables that better forecast indemnities (Appendix - Table A.1).

The analysis of the frequency of selection suggests that Boosting ensures a more stable set of selected variables than EN and LASSO. This result aligns with the theory: a more selective ML tool has a lower probability of being stable. Indeed, LASSO has a lower frequency of selection than EN and Boosting. In these latter models, even the most selected variables have a frequency of selection lower than 50%. On the other hand, EN is less selective than LASSO showing a higher frequency of selection. Boosting has ten variables constantly selected among many others, with a very high selection frequency compared to EN and LASSO.

The variable selection results allow identifying the variables that more influence the indemnity values (see appendix for further details). These are the average Value Added and its standard deviation (AVG_VA, sd_VA), the circulating capital standard deviation (K_CIRC_sd), the level and standard deviation of short-term liquidity (L_IMM_L1 and L_IMM_sd), the standard deviation of the total farm revenues (TR_sd), and labour costs (LAB_COS_L1). These results suggest that the core of the indemnities prediction relies on economic or financial aspects.

### 4.1.3  Trade-off between the number of variables and goodness-of-fit

A trade-off exists between the number of variables and goodness-of-fit: generally, the latter can be improved by increasing the number of regressors used. However, the results of our analysis suggest that a high goodness-of-fit does not always follow from a high number of regressors (i.e., much information) (Figure 4). Moreover, this result is quite counterintuitive based on the classical inference where a high number of independent variables generally corresponds to a low value of RMSE.

By assessing the pattern's collinearity properties, it is possible to determine which factors are essential for predicting indemnities and which ones are not; roughly half of the regressors used in the GLM show VIF > 5 (Fox *et al.*, 1992; Fox and Weisberg, 2012) demonstrating that the model is over defined.[9]

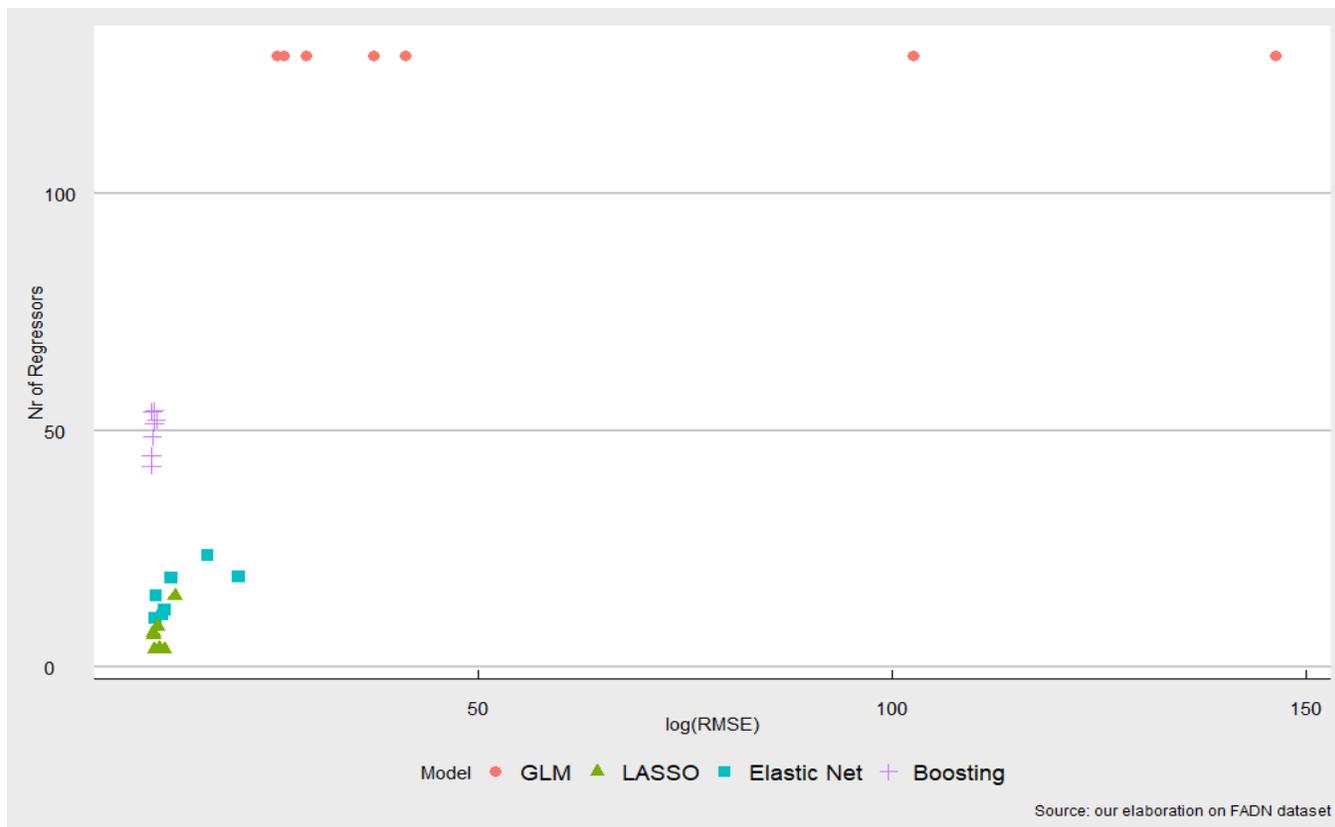

### 4.1.4

Figure 4 - Comparison of different models in the trade-off between the number of selected regressors and goodness-of-fit (log(RMSE)) in different years.

## 4.2  Economic assessment

As presented in the methodological section, the economic assessment considers the "affordability" and the "fairness "of the premiums, the multiannual balance of the mutual fund, and the fluctuation of its annual balances in the considered years.

### 4.2.1  Affordability

The affordability of the IST premium is assessed based on the four thresholds of relative premium discussed in the methodology session. However, we report only the more restrictive case of the 50th percentile threshold for the sake of brevity.

In this case, less than 5% of farms have a relative IST premium within the 50th percentile of the distribution of the ratio premiums paid over farm value-added. But, again, this is in line with similar studies (Goodwin and Mahul, 2004).

---

[9]This analysis is available upon request.

Figure 5 suggests that not considering the affordability of the premium could lead to a relevant overestimation of the degree of farmers' participation in the new insurance scheme.

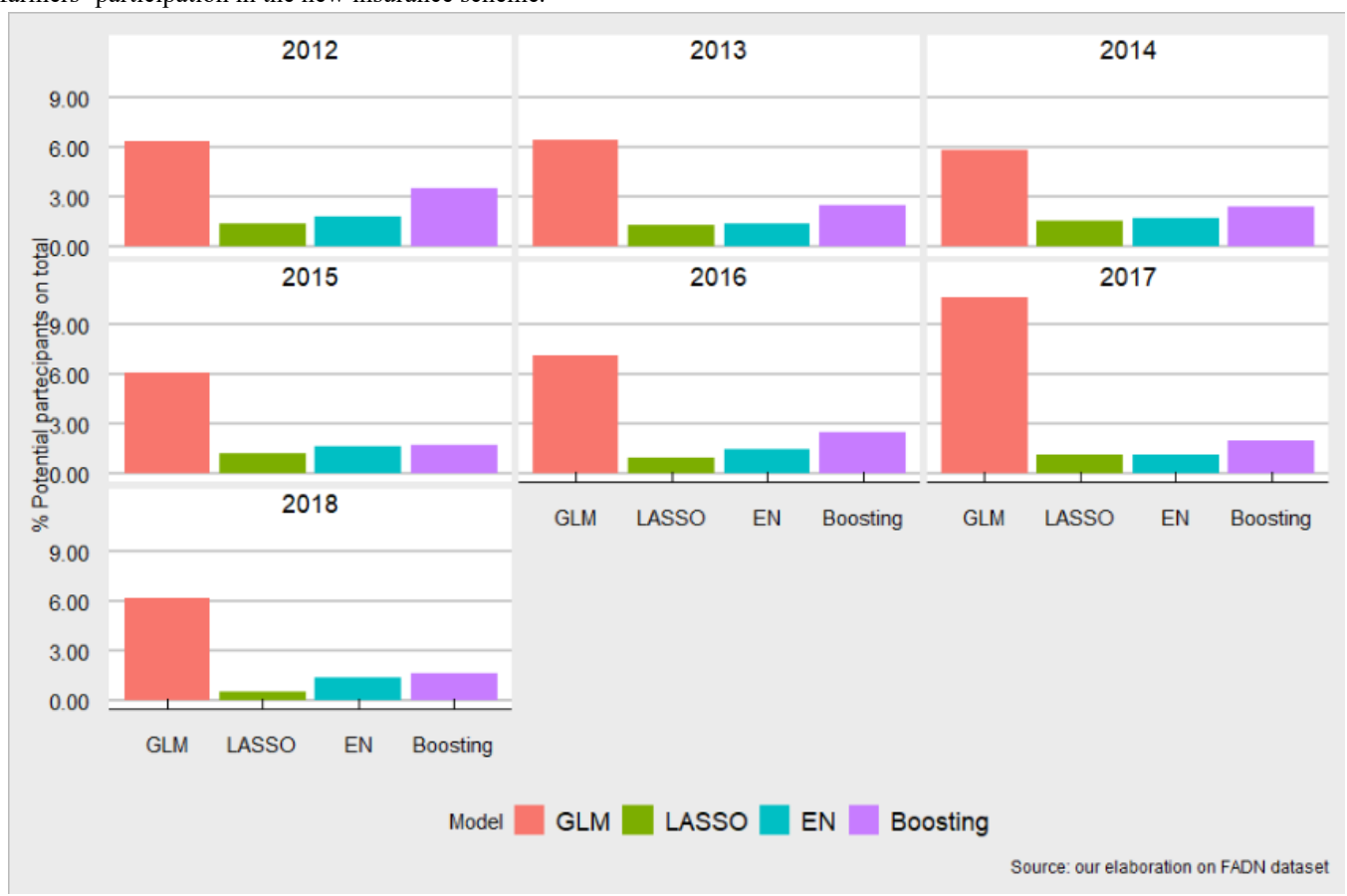

**Figure 5 - Potential participants to the IST according to the affordability criteria. Share of participants on the total sample in all considered years (2012-2018) and models - 50th percentile threshold**

Furthermore, when the 50th percentile distribution of the ratio premiums paid over farm value-added is considered, the four models show different results regarding premium affordability. Boosting allows having a larger share of farmers within the affordability constraint than the other ML tools, suggesting a better affordability performance and ensuring a higher degree of participation. Note that the use of GLM results could provide, in theory, an even larger share of farms facing affordable premiums. However, as will be shown shortly, this is because the resulting ratemaking, in this case, is not economically sustainable: the premiums, on average, are too low to ensure a sound economic balance for the mutual fund.

### 4.2.2 Fairness

To evaluate the fairness of the insurance design, we have compared the cases with the positive sum of net premiums (premium paid minus indemnity received) with the negative ones (Table VII). If we have high values for both figures, the premium is not in line with expected indemnities, and the ratemaking is unsatisfactory. This aspect is particularly detrimental for the negative values, i.e., the policyholders with underestimated risk. Indeed, considering the no-mandatory assumption, the subject that pays less than expected indemnity quickly leaves the scheme, with the effect that the only subjects that subscribe to the insurance in the next year will be the riskier.

This consequence is reflected in the balance sheet: the subjects with a positive net premium, which represents income for insurance, will decrease in number, while the policyholders with a negative net premium, which negatively affects the balance, will remain. This undermines the financial sustainability of the IST.

| Model | Sum of Net Premium>0 | Sum of Net Premium<0 | Sum of Net Premiums |
|---|---|---|---|
| GLM | 1'151'477 | -4'472'768 | -3'321'291 |
| LASSO | 981'324 | -909'701 | 71'623 |
| EN | 1'009'316 | -898'786 | 110'530 |
| Boosting | 1'184'908 | -1'184'485 | 423 |

**Table VII- Fairness level: differences between negative and positive net premium (50th percentile threshold)**

The result confirms that ML performs better than the GLM model. GLM suffers from a very negative amount of negative net premium that can threaten IST's survival in the following years. Furthermore, the sum of the net premium demonstrates the poor performance of GLM, which may have solvency problems. Among the ML tools, Boosting outperforms LASSO and EN ensuring a small gap between negative and positive net premiums.

### 4.2.3 Multiannual and annual balance sheet

The effect of using the estimation results on the economic sustainability of the scheme can be assessed by considering the level of the multiannual balance of the mutual fund (i.e., the sum of the premiums minus indemnities over the whole period). Using the results of the GLM for defining the premiums yields a significant negative economic outcome: the overall amount of indemnities exceeds that of the received premiums (Figure 6).

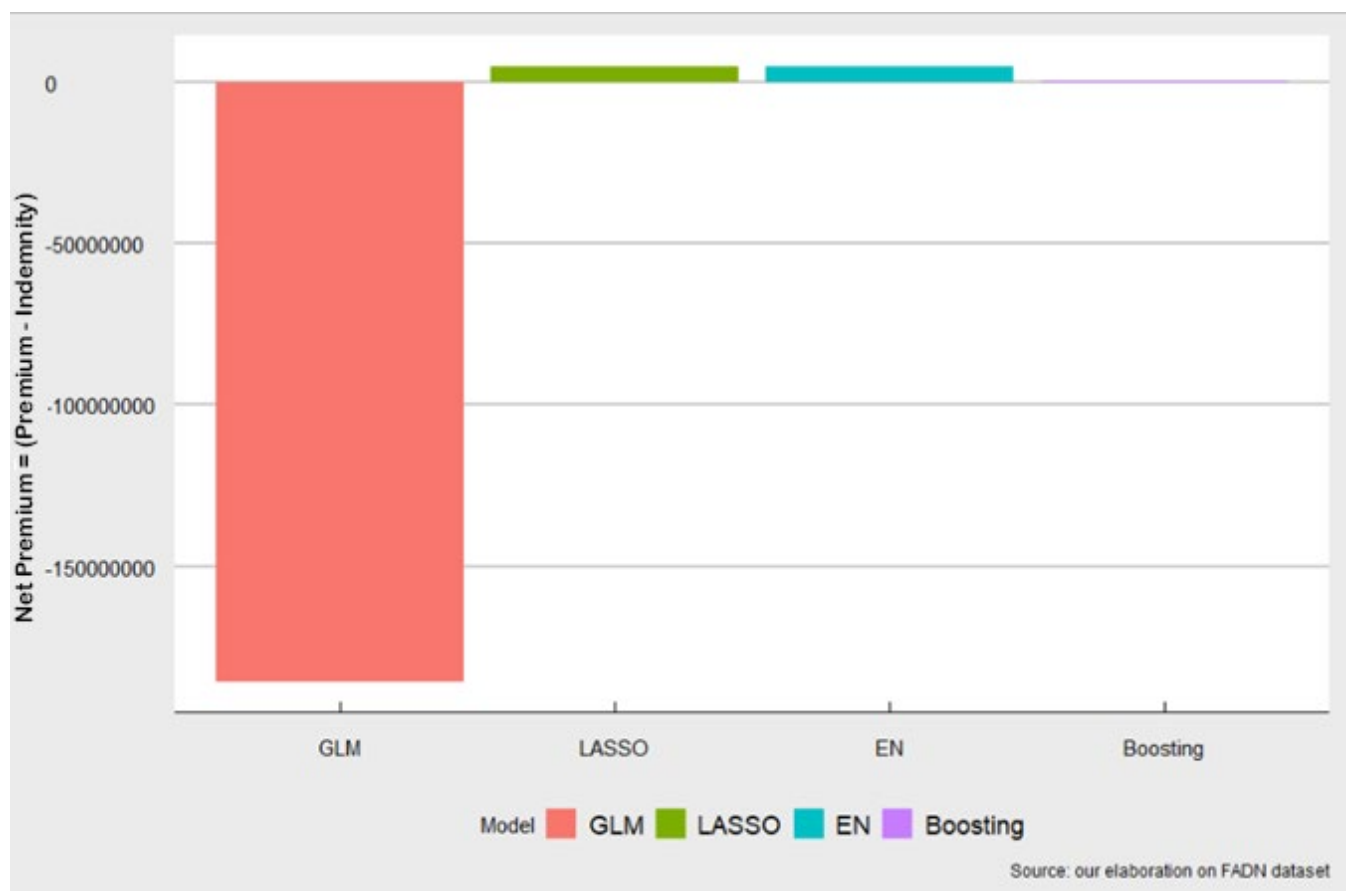

**Figure 6 - Multiannual (2012-2018) balance of the mutual fund (total net premium = sum of premiums-indemnities) as forecasted using the results of the four considered models in the ratemaking – ( 50th percentile threshold).**

The multiannual balance is more satisfactory when using the estimations of the ML models. Again, the LASSO and EN perform better than Boosting in this regard. These results suggested that ML, particularly EN and LASSO, can ensure the financial sustainability even if results slightly changed when applying less binding affordability constraints.

Finally, it is also essential to consider the variability of the annual balances over time to identify whether they reach a sizeable negative level in some years. This information helps the mutual fund set a high enough level of precautionary capital.

MF provides the annual balances, considering the tighter compatibility level for the four models (Figure 7).

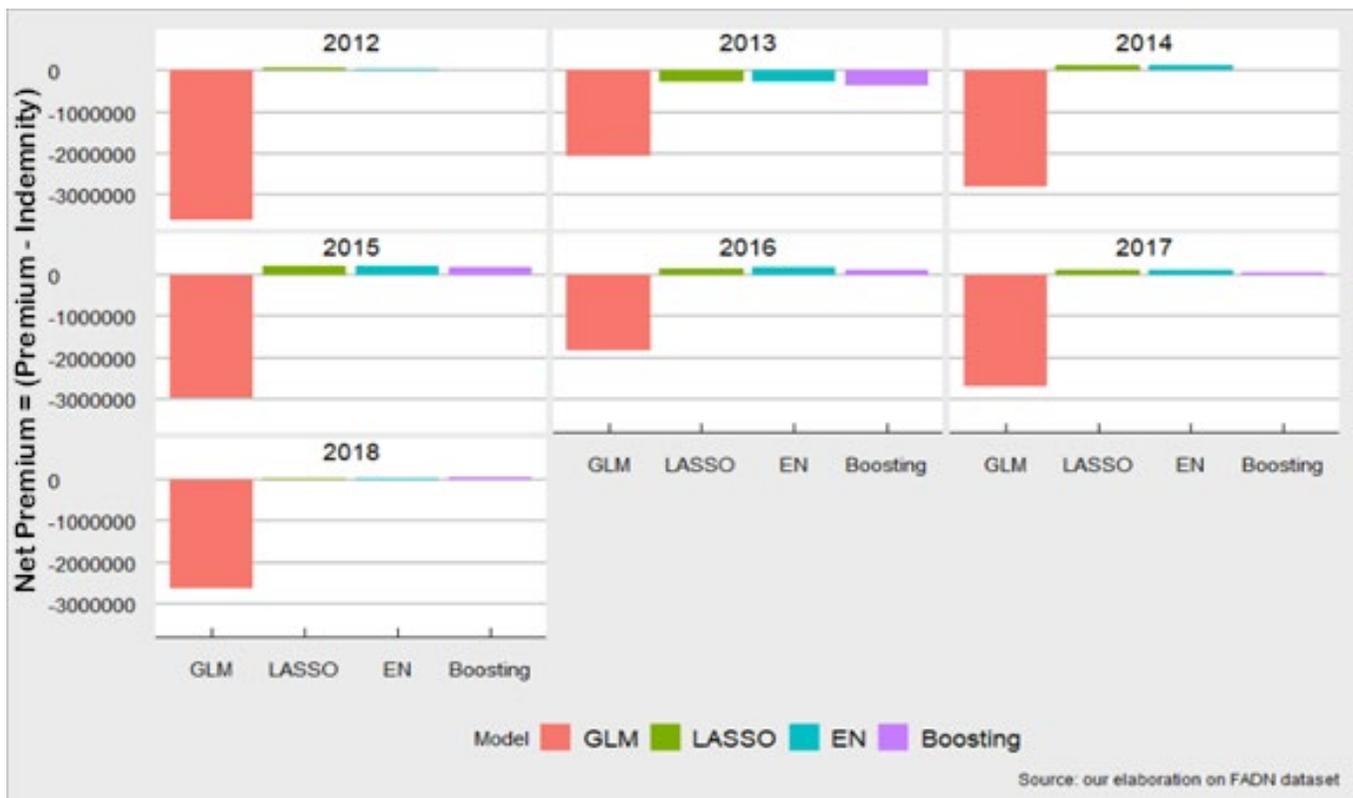

**Figure 7 - Variability over time of the annual balances of the Mutual Fund considering the four models.**

These results confirm the low economic sustainability of ratemaking when the GLM model results are used as the basis for ratemaking. In contrast, the three ML tools provide satisfactory results regarding the annual evolution of the balances. All three ML tools deliver results that ensure a relatively limited interannual variability of the balance. The worst year was 2013 if the balance sheet was not very harmful in all three ML tools compared to GLM for Mutual Funds.

This result aligns with other analyses developed in the USA for similar instruments (Goodwin and Mahul, 2004). However, some of the considered models, noticeably the Boosting, performed better in this regard.

# 5. Conclusions

This paper explored the pros and cons of using three ML tools that perform variable selection (LASSO, Elastic Net, and Boosting) to predict expected indemnities to inform insurance ratemaking, i.e., for establishing the levels of premiums. In addition, the econometric and economic performances of these ML tools were compared to the classical GLM models.

The results of this analysis are significant because ML tools, in particular Boosting, have some desired characteristics. ML models overcome the level of goodness-of-fit of GLM despite using fewer variables. Furthermore, the out-of-sample procedure with the year-to-year approach demonstrated the stability of the results of the ML estimates. ML seems to ensure good economic performance, too. The estimates obtained by ML allow a ratemaking that avoids short-term (annual) and long-term (multiannual) strong negative balance sheets, with the risk of failure for the institution managing the new scheme.

Our findings are of great importance to policymakers, insurers, and policyholders, pointing to the superiority of ML tools to the GLM models, which are currently extensively implemented in insurance design methods. In our study, using the considered ML tools allows the insurer to design fair and less expensive premiums thanks to decreasing the expense of gathering information (more variable selection) and a more precise forecast of indemnities. As a result, the policyholders pay a fair premium in line with their levels of risk. These aspects are paramount, especially when implementing new insurance schemes.

The results confirm that ML models will likely be more applied to agricultural economic issues in the coming years, as suggested by Jia *et al.*, (2019) and Storm, Baylis and Heckelei (2020). Furthermore, our analysis suggests this will also be the case in agricultural insurance.

The analysis has some limitations that should be considered. First, as is always the case with ratemaking for new insurance products, it is impossible to study farmers' behaviour explicitly. Hence, we cannot account for their responses to the introduction of the proposed scheme. Second, the analysis heavily relies on the data from the previous period. However, the proposed approach can be easily extended to account for the history of the considered individual farms. Furthermore, the analysis does not account for farmers' risk attitudes because this piece of evidence is currently not available. Finally, the lack of large enough, reliable data can constrain the capacity of mutual funds to use the proposed methodology.

# Supplementary Materials

This document briefly introduces Machine Learning (ML), including current developments and critical issues, and suggests references for further reading. The note is not intended to be exhaustive but only to give a preliminary introduction to the topic. Finally this document describes the ML tools used in the analysis and how the dummy variables are introduced in the models.

### A-1.        A BRIEF INTRODUCTION TO MACHINE LEARNING

Data Generation Processes (DGPs) are the basis of inference and can be defined as a process in which a set of input variables $x$, (independent variables) is associated with a function generating an output y (dependent variable). Econometricians usually use a stochastic DGP or $y = f(x, \beta; \varepsilon)$ where output $y$ deriving from a function $f(\cdot)$ of input o predictor variables $x$, parameter $\beta$, and random noise $\varepsilon$ (Hamilton, 1994; Greene, 2012). The analysis in economics is based on a theoretical framework, and its primary goal is to make inferences rather than obtain a large predictive capacity of the model (Breiman, 2001; Charpentier, Flachaire and Ly, 2018).

Machine Learning (ML), first defined by Samuel (1959), differs in this regard. Indeed, Iskhakov, Rust, and Schjerning (2020 page 2) define ML with an extensive formulation as: "*the scientific study of the algorithms and statistical models that computer systems use to perform a specific task without using explicit instructions and improve automatically through experience.*"

This new statistical toolbox has also generated a deep rift between classical statisticians, linked to inference, and those who saw in ML new possibilities. Breiman (2001) highlighted this division by defining the clash's terms. ML tools have different objectives than those used in statistical inference. First, the objective of ML is the ability to predict concerning inference (many tools are biased by definition). Second, ML has more ability to map or reduce the dimensionality of problems. Third, ML can not establish a DGP based on a theoretical framework. ML can deduce the DGP from data (data-driven). Allowing the discovery of new DGPs that may not have been previously defined in the literature or the theoretical context (Breiman, 2001). Last but not least, ML is very efficient in the use of computational resources (Charpentier, Flachaire and Ly, 2018; Athey and Imbens, 2019; Efron, 2020).

ML has come under criticism regarding different aspects. First is the lack of mathematical models used at the base of inference: this concept has been defined by Cox (2001) as "*Abandoning mathematical models comes close to abandoning the historic scientific goal of understanding nature*" and also supported by (Parzen, 2001; Efron, 2020). Second, ML is usually not a BLUE estimator but only a Best Linear (BLE) estimator because it is generally biased (van de Geer, 2016). Third, Breiman (2001) introduced some additional issues for ML: i) the possibility of discovering multiple DGPs (defined as Rashomon's effect); ii) the necessity to reduce the complexity of the model or to consider the trade-off between accuracy and interpretability of model (called Occam's effect); iii) the need to increase the number of variables to improve accuracy. The latter increases the model's complexity and estimation and the inability to find a precise solution to complex problems. This is called by Breiman (2001) "curse of dimensionality" or Bellman's effect (Bellman (1957)). The "curse of dimensionality" could be decreased using ML, which allows selecting only the necessary variables to estimate the DGP. Another issue is whether a very selective ML tool should prefer or uses more regressors to be more precise. There is no single answer because it depends on the problem at stake. However, it is always helpful to obtain a satisfactory compromise between complexity and interpretability (Efron, 2020).

Another issue with ML is the lack of confidence intervals (that is a point of strength in econometric models) and the possibility of using the marginal effects that are useful to describe the relationship between the dependent and independent variable (Leeb and Pötscher, 2006b, 2006a; Lee *et al.*, 2016; Taylor and Tibshirani, 2018). This problem is common to all algorithms used to select the variables (Leeb and Pötscher (2006a, 2006b)). However, according to Mullainathan and Spiess (2017) and Athey and Imbens (2019), confidence intervals are functional to a specific scope. In particular, the trade-off between accuracy (higher in ML by construction) and confidence interval analysis should be carefully considered. At the same time, the researcher has to reflect on whether some properties may or may not be relevant (Mullainathan and Spiess, 2017; Athey and Imbens, 2019). For example, when many potential regressors are available, depending on the research questions at stake, the high level of collinearity may be considered more critical than not having confidence intervals. This trade-off will interest the researchers because a valuable and trustworthy substitute for confidence intervals in ML has not been proposed yet (Lu *et al.* 2017; Liu, Markovic, and Tibshirani 2018; Zhang *et al.* 2020; Zrnic and Jordan 2020). Furthermore, according to Leeb and Pöscher (2006) and Leeb and Pötscher (2006, 2008), making inferences after variable selection is difficult. In particular, the bootstrap can yield peculiar results (Tibshirani *et al.*, 2016)). To overcome this problem, Lee *et al.* (2016) and Tibshirani *et al.* (2016) proposed to use a conditional approach leading to a truncated normal reference distribution.

Furthermore, Mullainathan and Spiess (2017: Figure 2 page 97) used the comparison between multiple DGPs to obtain a comparison similar to that found in econometric inference for confidence intervals. Analyzing the frequency with which a variable is selected in all simulations, it is possible to obtain the "strength" of the variables in establishing the DGP. This procedure has been used in our study.

The ML approach is not widely used in economics (Athey and Imbens, 2019) mainly because economists base their analyses on inference than on the model's predictive capacity. As a result, the fracture reported in Breiman (2001) is no longer healed. As reported in recent works, this mending can only take place when statisticians and econometricians become aware that these tools are complementary and not opposed (Einav and Levin, 2014; Varian, 2014; Athey, 2017; Kleinberg *et al.*, 2017; Mullainathan and Spiess, 2017; Charpentier, Flachaire and Ly, 2018; Athey and Imbens, 2019; Rust, 2019; Efron, 2020; Iskhakov, Rust and Schjerning, 2020).

## A-2.  A BRIEF DESCRIPTION AND COMPARISON OF THE ML TOOLS USED IN THE ANALYSIS

LASSO and Elastic Net[10] are ML methods based on *regularisation* or *shrinkage* methodology (Hastie, Tibshirani and Friedman, 2009). These allow us to obtain threefold results: select variables, enhance prediction accuracy and reduce collinearity.

Shrinkage methodology has been used for the ridge regression or Tikhonov regularisation: while the Least Squared seeks to minimize the sum of squared residual $\|Ax - b\|_2^2$ the ridge regression adds a regularisation term $\|\Gamma x\|_2^2$. The objective of the ridge regression (Loss function) is to minimize the Mean Square Error (MSE) of $\|Ax - b\|_2^2 + \|\Gamma x\|_2^2$ (with $\|\Gamma x\|_2^2$ is Euclidean norm defined as $l_2$).

The LASSO regression uses the following modification of the Tikonov term: $\|Ax - b\|_2^2 + |\Gamma x|$ (with $|\Gamma x|$ also defined as $l_1$). Elastic Net uses an intermediated value of the regularisation term that lies between Euclidean norm and Absolute norm (between $l_1$ and $l_2$).

We can rewrite Loss Function as minimization of MSE of i) RIDGE $= L_{\text{Ridge}} = RSS + \lambda \sum_{j=1}^{p} \beta_j^2$; ii) LASSO$= L_{\text{Lasso}} = RSS + \lambda \sum_{j=1}^{p} |\beta_j|$; iii) EN is given with a combination of Ridge and LASSO thanks to a coefficient $\alpha : L_{\text{Elastic Net}} = \frac{RSS}{2n} + \lambda \left( \frac{1-\alpha}{2} \sum_{j=1}^{p} \beta_j^2 + \frac{\alpha}{2} \lambda \sum_{j=1}^{p} |\beta_j| \right)$.

$\lambda$ denotes a tuning parameter indicating the strength of the penalty term, and it is the hyper-parameter of the Shrinkage

---

[10] Elastic Net regression is part of the Shrinkage Regression Family (Tibshirani, 1996). This, unlike the classical statistical prediction, aims to find a function that gives a "good" prediction of $y$ as a function of $x$ where "good" means it minimizes or maximize some objective of the inference as RSS, Deviance, AUC, AIC, BIC (Varian, 2014).

Tool. $\lambda$ is set via Cross-Validation[11].

Boosting is a machine learning tool that primarily aims to reduce distortion and variance by using a "weighting" methodology that transforms weak predictors into strong predictors. The central hypothesis of "boosting" is the division into strong and weak regressors. While the former provides a high contribution to the explanation of the model, the latter is not very important for prediction. Boosting methodology weights the regressors differently based on their explanatory power. The ability to convert a mediocre regression into one that works exceptionally well through a learning algorithm is one of the strengths of Boosting[12].

The algorithm performs many simulations to obtain the correct weight. Each of these simulations is a basis for the next phase (learning method). For example, the first step uses a constant weight for each regressor, but only a few have high predictive power.

There are several implementations of the Boosting approach. We use the Gradient Boosting algorithm (Friedman, 2001). The Loss Function in Boosting is defined as $\hat{t}(x) = argmin \sum_{i=1}^{N} L(\hat{y}_i, y_i)$, with $\hat{y}_i$ = predicted value, $y_i$ the observed value and i is the observations, and $L$ is a function that we can also use to obtain MSE, RMSE, or Huber (Adaptative) Loss, Entropy, or Exponential Loss. The meta-algorithm used in Gradient Boosting is the following:

- Establish the number of iterations $M = \{1, \ldots, m\}$ and the learning rate, or shrinkage factor, $\nu$ ($M$ and $\nu$ are the hyperparameters to set by the statistician)
- In the beginning, one starts with a causal tree and does the first stage, where the results $F_0 = L_0(\hat{y}_i, y_i)$.
- In the second stage, one takes $y_i - F_0(X_i) = h_0(x)$ with $F_0(X_i) = \hat{y}_i$
- Uses the RMSE of the first stage $L_{RMSE} = \sqrt{\frac{1}{2}\left(y_i - F_0(X_i)\right)_i^2}$ and use the negative gradient $r_{i,m}(x) = -\left[\frac{\partial L_{RMSE}}{\partial F(X_i)}\right]_{F(X_i) = F_{m-1}(X)}$ to find the so-called pseudo-residuals.
- Fits the leaner $h_m(x)$ using the training-set $h_m(x) = \left\{\left(x_i, r_{i,m}\right)\right\}_{i=1}^{n}$
- Computes multiplier $\gamma_m$ to solve the problem

$$\gamma_m = \underset{\gamma}{argmin} \sum_{i=1}^{n} L\left(y_i, F_{m-1}(x_i) + \nu\,\gamma_m\,h_m(x_i)\right)$$

where $h_m(x_i)$ represents a weak learner of fixed depth, $\gamma_m$ is the step length and $\nu$ is the learning rate or shrinkage factor
- The model $F_m(x) = F_{m-1}(x) + \gamma_m h_m(x)$ is updated
- In the end, the final output $F_M(x)$ is obtained

The comparison between Boosting and shrinkage methods, especially in economics, deserves reflection. Boosting allows for very high accuracy without needing a mathematical model definition. Conversely, Elastic Net and LASSO, one must first define the functional form of the regressors.

The choice of one or the other approach depends on assumptions and research questions. For example, if having high accuracy without first setting the functional shape of the regressors is important, Boosting certainly is the best choice. On the other hand, LASSO and EN should be preferred if it is necessary to verify a more structural model based on economic theory, assumptions, and constraints.

## A-3.   COMPARISON OF THE CONSIDERED ML APPROACHES

According to our empirical results, we can compare the considered ML approaches and the GLM considering several characteristics. A synthesis is provided in the following Figure.

---

| Characteristic | GLM | LASSO | EN | Boosting |
|---|---|---|---|---|
| Accuracy | Poor | Fair | Fair | Good |
| Interpretability | Good | Good | Good | Fair |
| Parsimony | Poor | Good | Good | Fair |
| Stability of selected variables | Good | Poor | Poor | Fair |
| Handling of data with no-predetermined mathematical shape | Poor | Poor | Poor | Good |
| Treatment of Multicollinearity | Poor | Good | Good | Good |
| Automatic (requires little tuning) | Good | Good | Fair | Fair |

| Legend | |
|---|---|
| ▲ | Good |
| ◆ | Fair |
| ▼ | Poor |

**Figure 1 - Comparison of some characteristics of the considered ML tools**

**Accuracy** (or Performance - goodness-of-fit): GLM has low performance (caused by high collinearity and overfitting effect (Fan and Lv, 2010)). LASSO and EN obtain adequate performance, while Boosting achieves better results than others.

**Interpretability**: According to Erasmus, Brunet and Fisher (2020), interpretability and understandability are strictly linked. However, a more naïve model is more understandable by a larger audience. For example, the results of GLM can be interpreted as those obtained by LASSO and EN. Conversely, Boosting makes interpreting the results harder because of its peculiarity. These regressors have a mathematical shape that is not directly interpretable.

**Parsimony:** refers to the ability to reduce the number of variables selected: LASSO is better than EN and Boosting in this regard.

**Stability of selected variables:** Boosting has high stability (also called consistency) in comparison with EN and particularly with LASSO (Fan and Lv (2010)).

**Handling of data without imposing a predetermined mathematical shape**. For GLM, LASSO, and EN, the variables must be introduced with a predetermined mathematical formulation such as logarithm, squared root, or exponential. Conversely, in Boosting, this transformation is not required. Hence, it can select different types of shapes of regressors, allowing a better description of the real DGP

**Treatment of multicollinearity:** with the high-dimensional setting (i.e., the high number of regressors), this issue becomes very dramatic, generating negative consequences in predicting the output (James et al., 2013). According to Mason and Brown (1975), penalized regression can reduce the multicollinearity problem significantly. Hence LASSO and EN are expected to reduce multicollinearity (Zou and Hastie, 2005). According to our results, all considered ML approaches strongly reduce multicollinearity.

**Automatic setting:** Differently from GLM, the setting in ML is essential. As explained earlier in the document, it is necessary to set the hyperparameters of the different ML approaches. (e.g., $\lambda$ for LASSO, $\lambda$, and $\alpha$ for EN, number of iterations $M$, and learning rate $\nu$ for Boosting). Accordingly, LASSO needs a more specific setting than EN. Finally, Boosting requires the setting of a higher number of parameters.

To conclude, the results of our analysis suggest that, in the considered empirical case, ML overcame the trade-offs between Accuracy and Interpretability and between Parsimony and stability of selection. Furthermore, if using a limited number of variables is the priority (e.g., because of high information costs), LASSO and EN perform better than Boosting. However, these are characterized by the lower stability of the selected variables. In contrast, this is not the case for Boosting. In our case, it performs better and has a more stable set of selected variables but lower interpretability than EN and LASSO.

## A-4.     TREATMENT OF DUMMIES WITH MULTIPLE CATEGORIES: THE GROUPED APPROACH

The grouped regression has been used to handle the categorical regressors assuming multiple discrete values (i.e., levels). To explain this issue, let us take the case of the five Italian macro-regions: North-West (NOR), North-East (NOC), Center (CEN), South (MER), and Islands (INS). We use one of the categories as contrast (for example, NOR), assigning to the other values 0 or 1. Without variable selection, we can compare all other categories with NOR. In contrast, in the case of variable selection, we may have that NOC is not selected. In this case, the contrast is no longer NOR but NOR + NOC, which does not allow a clear interpretation of the model. To avoid this issue, we have imposed a selection of all or none of the categories on the algorithm. In practice, we "group" these categories and create a constraint to selecting the "group" and not a single class. This method is called "Grouped Regression" (Bühlmann and van de Geer, 2011; Qian, Yang and Zou, 2016) and is applied to all categorical variables with classes higher than two.

## 6. LIST OF VARIABLES

This table report all the variables with the percentage of selection in the machine learning model. The code refers at the variables of table II with three tpyes of suffixes: AVG is average of the past three year ( from $t-1$ to $t-3$), sd is the standard deviation referred to past three year ( from $t-1$ to $t-3$) and L1 is the value at the $t-1$.

| Description | Code |
|---|---|
| *Farm characteristics* | |
| Utilized agricultural land [Ha] | LND |
| Livestock Units [LU] | LU |
| Labour input [AWU] | LAB |
| Total revenues [€] | REV |
| *Farm production characteristics:* | |
| Livestock intensity [LU/Ha] | LU_I |
| Machinery intensity [kWh/Ha] | MACHIN |
| Labor per UAA [UL/ha] | LAB |
| Land productivity [€/ha] | VA |
| Specialization [Herfindhal Index] | H Index |
| Other Gainful Activities [%] | OGA |

| Description | Code |
|---|---|
| ***Farm financial characteristics:*** | |
| Fixed cost [€/ha] | FXCOST |
| Current over total costs [%] | CURCOST |
| Labor over total costs [%] | LBRCOST |
| Insurance premia over total costs [%] | INSURE |
| Relative amount of fixed capital [%] | FXK |
| Relative amount of debts [%] | DEBT |
| Relative amount of net worth [%] | NETK |
| ***Farm policies* (CAP)** | |
| Decoupled Direct Payments [€] | DDP |
| Coupled Direct Payments [€] | CDP |
| Rural Development Policies- Agroenvironmental [€] | RDP_AES |
| Rural Development Policies - Less Favoured Area [€] | RDP_LFA |
| Rural Development Policies for Investments [€] | RDP_INV |
| ***Other farm characteristics as number of observations:*** | |
| Number of farm | N_FARMS |
| Sole proprietor farm | INDIV |
| Gender of the holder | MALE |
| Young holder | YOUNG |
| Organic farms | ORGAN |
| Plain regions | PLAIN |
| Hill regions | HILL |
| Mountain regions | MOUNT |